\documentclass[10pt]{article}
\usepackage[margin=1in]{geometry}
\usepackage{amsmath,amsfonts,amssymb}
\usepackage{graphicx}
\usepackage{bm}
\usepackage[affil-it]{authblk}
\usepackage{natbib}
\usepackage[bf,font={small,sl}]{caption}
\usepackage[pdftex,dvipsnames]{xcolor}
\usepackage[colorinlistoftodos, textsize=small]{todonotes}
\presetkeys{todonotes}{fancyline, backgroundcolor=Apricot, bordercolor=white, textcolor=BlueViolet}{}

\usepackage{parskip}
\usepackage{booktabs}
\usepackage{url}
\usepackage{soul}

\newcommand{\x}{\bm{x}}
\newcommand{\z}{\bm{z}}
\newcommand{\bbeta}{\bm\beta}

\title{\bf Understanding step selection analysis\\ through numerical integration}
\author{Théo Michelot$^{1}$\footnote{Email: \texttt{theo.michelot@dal.ca}}, Natasha J.\ Klappstein$^{1}$, Jonathan R.\ Potts$^{2}$, John Fieberg$^3$}
\affil{\normalsize$^1$Dalhousie University, $^2$University of Sheffield, $^3$University of Minnesota}
\date{}

\begin{document}
\maketitle

\begin{abstract}
  Step selection functions (SSFs) are flexible statistical models used to jointly describe animals' movement and habitat preferences. The popularity of SSFs has grown rapidly, and various extensions have been developed to increase their utility, including the ability to use multiple statistical distributions to describe movement constraints,  interactions to allow movements to depend on local environmental features, and random effects and latent states to account for within- and among-individual variability. Although the SSF is a relatively simple statistical model, its presentation has not been consistent in the literature, leading to confusion about model flexibility and interpretation. 
  We believe that part of the confusion has arisen from the conflation of the SSF model with the methods used for statistical inference, and in particular, parameter estimation. Notably, conditional logistic regression can be used to fit SSFs in exponential form, and this model fitting approach is often presented interchangeably with the actual model (the SSF itself). However, reliance on conditional logistic regression reduces model flexibility, and suggests a misleading interpretation of step selection analysis as being equivalent to a case-control study. 
  In this review, we explicitly distinguish between model formulation and inference technique, presenting a coherent framework to fit SSFs based on numerical integration and maximum likelihood estimation. We provide an overview of common numerical integration techniques (including Monte Carlo integration, importance sampling, and quadrature), and explain how they relate to popular methods used in step selection analyses. 
  This general framework unifies different model fitting techniques for SSFs, and opens the way for improved inferential methods. In this approach, it is straightforward to model movement with distributions outside the exponential family, and to apply different SSF model formulations to the same data set and compare them with AIC. By separating the model formulation from the inference technique, we hope to clarify many important concepts in step selection analysis.
\end{abstract}

\section{Introduction}

\begin{figure}[htbp]
    \centering
    \includegraphics[width=0.8\textwidth]{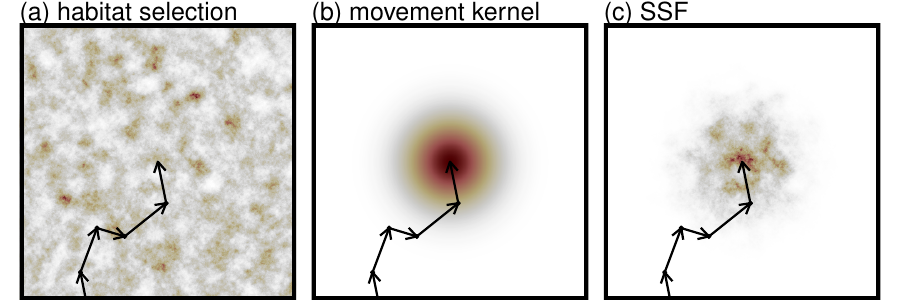}
    \caption{Example model components at some time $t$, where the last point $\x_t$ is at the centre of each panel and the previous steps are shown as black segments: (a) habitat selection function $w(\x_t, \z)$, (b) movement kernel $\phi(\z \vert \x_t)$ based on distance from $\x_t$, and (c) resulting step selection function $w(\x_t, \z) \phi(\z \vert \x_t)$. The integral on the denominator of Equation \ref{eqn:ssf} is the volume under the step selection function, which is required to transform the step selection function into a probability distribution (sometimes called the step density).}
    \label{fig:ssf}
\end{figure}

The increased availability of animal tracking data has led to the widespread use of statistical methods to estimate habitat selection at the scale of the observed movement step. Perhaps the most common model is the step selection function \citep[SSF;][]{rhodes2005, fortin2005}, whereby the likelihood of moving to the spatial location $\x_{t+1}$ given previous locations $\x_{1:t} = \{ \x_1, \x_2, \dots, \x_t \}$ is in the following form,
\begin{equation}
  \label{eqn:ssf}
  p(\x_{t+1} \mid \x_{1:t}) = \frac{w(\x_t, \x_{t+1}) \phi(\x_{t+1} \mid \x_{1:t})}{\int_\Omega w(\x_t, \z) \phi(\z \mid \x_{1:t}) d\z},
\end{equation}
where $\Omega$ is the study region. The function $w$ describes the effects of environmental variables \citep[e.g., resources, risks, and environmental conditions;][]{matthiopoulos2023}, and $\phi$ accounts for the effects of movement constraints (e.g., on the range of observed step lengths).  The habitat selection function is often assumed to take an exponential (or ``log-linear'') form, i.e., $w(\x_t, \x_{t+1}) = \exp \{ h(\x_t, \x_{t+1}) \bbeta^\intercal_h \}$, where $h(\x_t, \x_{t+1})$ is a vector of habitat variables for the step, and $\bbeta_h$ is the vector of associated selection parameters. The form of $\phi$ reflects assumptions about movement patterns of the animal, and it is often written as a function of the step length and turning angle to capture movement speed and tortuosity of an animal's movement. We call ``step selection function'' the numerator of Equation \ref{eqn:ssf}, but the terminology is not consistent across the literature, and the term has been used variously to refer to $w$, to $w \times \phi$, or to the whole right-hand side of Equation \ref{eqn:ssf}. We choose to define $w \times \phi$ as the SSF to reflect the fact that an animal's selection of a step is based on both habitat preferences and movement constraints. Figure \ref{fig:ssf} shows an example step selection function with the two model components.

Step selection analysis refers to a wide range of methods for applying SSFs to animal tracking data, with the aim to estimate the parameters of the habitat selection function $w$ and the movement kernel $\phi$. Although the data-generating mechanism for this model is described by Equation \ref{eqn:ssf}, this is a difficult statistical problem due to the presence of the integral in the denominator. This integral is required in Equation \ref{eqn:ssf} to ensure that $p(\x_{t+1} \mid \x_{1:t})$ integrates to 1, i.e., that it is a valid probability density function with respect to $\x_{t+1}$ \citep{rhodes2005, forester2009}. It also has a more intuitive interpretation: to evaluate the likelihood of a given step, we weigh its suitability against the suitability of all other possible steps in the study region. Here, ``suitability'' refers both to the habitat quality of a location (as captured by $w$) and to its accessibility (as captured by $\phi$). This integral cannot generally be calculated analytically, because the integrand (i.e., the expression that is integrated) depends on $w$, which is usually a function of environmental covariates with no mathematically-convenient functional forms. That is, the integral cannot be rewritten in terms of simple functions that could be directly implemented with a computer, and so Equation \ref{eqn:ssf} cannot generally be evaluated for a given movement track and set of parameters.

Although the integral in Equation \ref{eqn:ssf} cannot be computed directly, methods have been developed to replace the expression by a tractable approximation. In some cases, such approximations are equivalent to applying conditional logistic regression (CLR) to a case-control data set, where each observed location (``case'') is associated with a set of locations from the landscape \citep[``controls'';][]{forester2009}. This has been a popular framework for step selection analysis, because CLR can be fitted quickly and conveniently using statistical software \citep[e.g., using the survival package in R;][]{therneau2023}. Consequently, SSFs are often conflated with CLR, even though the latter is merely a convenient tool to fit the former by approximating the likelihood in some special cases. In our view, this presentation can lead to confusion about model interpretation, and reduces the flexibility of step selection analyses. In particular,  we can avoid the need to make strong assumptions about the functional forms of $w$ and $\phi$ if we are willing to use numerical methods other than CLR for parameter estimation.

In this review, we show that most methods used to estimate parameters in step selection analyses can be viewed as applications of numerical integration. A variety of numerical integration techniques are commonly applied in statistics and mathematics to approximate integrals in cases where there is no known formula to compute them exactly. We will present several numerical tools developed for this purpose, and describe their utility in step selection analyses. This perspective suggests we can contrast existing methods (e.g., to identify those with lower approximation errors), and opens the way for improved inferential techniques in step selection analysis. Lastly, we hope that our review will motivate further exploration of numerical integration and estimation methods, which have broad utility across a wide range of ecological applications.

\section{Likelihood approximation in step selection analysis}
\label{sec:mle}

\subsection{Maximum likelihood estimation}

We consider $n$ observed locations $\x_1, \x_2, \dots, \x_n$, recorded without error at regular time intervals. The goal of  step selection analysis is to estimate the parameters $\bbeta_h$ of the habitat selection function $w$, which quantify the strength of selection or avoidance of spatial covariates, and the parameters $\bbeta_m$ of the movement kernel $\phi$, which quantify movement tendencies (e.g. speed and tortuosity). We denote as $\bbeta = (\bbeta_h, \bbeta_m)$ the vector of all parameters. Based on the standard assumptions that each step follows the model of Equation \ref{eqn:ssf}, and that successive steps are conditionally independent (i.e., given past locations), the likelihood of the parameters under the step selection model can be written as
\begin{equation*}
  L(\bbeta; \x_1, \dots, \x_n) =
  \prod_{t = 1}^{n-1} p(\x_{t+1} \mid \x_{1:t}) =
  \prod_{t = 1}^{n-1} \frac{w(\x_t, \x_{t+1}) \phi(\x_{t+1} \mid \x_{1:t})}
  {\int_\Omega w(\x_t, \z) \phi(\z \mid \x_{1:t}) d\z}.
\end{equation*}

Note that both $w$ and $\phi$ depend on some of the parameters in $\bbeta$ (specifically, $w$ depends on $\bbeta_h$ and $\phi$ on $\bbeta_m$), but we do not make this explicit for notational convenience. The likelihood (or log-likelihood) can be optimised numerically with respect to $\bbeta$, for example using the function \texttt{optim} or \texttt{nlm} in R, to obtain maximum likelihood estimates of $\bbeta$. For an overview of maximum likelihood estimation in an ecological context, see for example Chapters 6--7 of \cite{bolker2008} or Chapter 10 of \cite{fieberg2022}.

Model fitting requires computing the likelihood function for the observed data, and therefore evaluating the integral
\begin{equation}
  \label{eqn:integral}
  I = \int_\Omega w(\x_t, \z) \phi(\z \mid \x_{1:t}) d\z
\end{equation}
for each time step $t \in \{1, 2, \dots, n-1 \}$. In this section, we describe several methods to approximate $I$ by some quantity $\widehat{I}$, which can be substituted in the likelihood formula to carry out approximate inference. The main approaches are summarised in Table \ref{tab:summary}.

\begin{table}[htbp]
    \centering
    \begin{tabular}{lcccc}
    \toprule
         & Approximate likelihood & $\bm{z}_k \sim$ ? & References \\
         \midrule
         MC & $\frac{w(\x_t, \x_{t+1})}{\sum_{k = 0}^K w(\x_t, \z_k)}$ & $\phi(\cdot \mid \x_{1:t})$ & \cite{fortin2005} \\[5mm]
         UMC & $\frac{w(\x_t, \x_{t+1}) \phi(\x_{t+1} \mid \x_{1:t})}{\sum_{k = 0}^K w(\x_t, \z_k) \phi(\z_k \mid \x_{1:t})}$ & uniform & \cite{forester2009} \\[5mm]
         IS & $\frac{ w(\x_t, \x_{t+1}) \phi(\x_{t+1} \mid \x_{1:t}) }{ \sum_{k = 0}^K  w(\x_t, \z_k) \phi(\z_k \mid \x_{1:t}) / h(\z_k \mid \x_{1:t})}$ & user-defined $h$ & \cite{forester2009} \\[5mm]
         UQ & $\frac{ w(\x_t, \x_{t+1}) \phi(\x_{t+1} \mid \x_{1:t}) }{ \sum_{k = 0}^K w(\x_t, \z_k) \phi(\z_k \mid \x_{1:t})}$ & regular grid & \cite{rhodes2005} \\
    \bottomrule
    \end{tabular}
    \caption{Summary of most common numerical integration approaches used in step selection analyses: Monte Carlo with known movement kernel (MC), uniform Monte Carlo (UMC), importance sampling (IS), and uniform quadrature (UQ). The columns include the approximate likelihood of a step from $\x_t$ to $\x_{t+1}$, method for determining the distribution of integration points $\z_k$, and references to key papers presenting each approach.}
    \label{tab:summary}
\end{table}

\subsection{Monte Carlo integration}
\label{sec:mc}

We use the term ``Monte Carlo integration'' to refer to all forms of numerical integration that rely on random sampling (i.e., all methods presented in this review, except for quadrature). Monte Carlo integration is a method for evaluating an integral of the form $\int_\Omega f(\z) g(\z) d\z$, where $f$ is a probability density function \citep[Section 3.2 of][]{robert2010}. The general idea is to generate a sample from the distribution $f$, and use it to find an unbiased estimate of the integral. It can be shown that
\begin{equation}
  \label{eqn:mc1}
  \int_\Omega f(\z) g(\z) d\z \approx \frac{1}{K} \sum_{k = 1}^K g(\z_k),\quad \text{where } \z_k \sim f,
\end{equation}
with the error of the approximation decreasing as $K$ increases.  Throughout this review, we use the notation ``$\z_k \sim f$'' to indicate that $\z_k$ follows the distribution with probability density function $f$. 

The most common approaches to fitting SSFs can be viewed as different forms of Monte Carlo integration applied to the integral in Equation \ref{eqn:integral}, which result from different choices of the functions $f$ and $g$ (always with the constraint that $f \times g = w \times \phi$). Generally, this choice may impact the accuracy and precision of the approximation, so choosing it requires thought \citep[see Section \ref{sec:imp} and][]{rizzo2019}.

Maximum likelihood estimation based on the numerical approximation of the integral in Equation \ref{eqn:mc1} defines a general framework of approximate inference for SSFs. Methods of inference based on Monte Carlo likelihood approximations are common in econometrics, where they are usually called \emph{simulated} maximum likelihood estimation, to highlight their inherent stochasticity \citep[Section 3.1.2 of][]{gourieroux1996}. Even though the Monte Carlo estimator of the integral given in Equation \ref{eqn:mc1} is not biased, simulated maximum likelihood estimators generally are, due to the log-transformation of the likelihood before it is maximised \citep{gourieroux1996}. As a result, Monte Carlo-based step selection parameter estimators are biased, but this bias decreases with the number of integration points. 

Using Monte Carlo integration, Equation \ref{eqn:integral} can be approximated as $I \approx \frac{1}{K} \sum_{k=1}^K g(\z_k)$, for functions $f$ and $g$ chosen such that $\z_k \sim f$ and $f \times g = w \times \phi$. In this review, we include the observed location $\x_{t+1}$ as an additional integration point in the random sample $\{ \z_1, \dots, \z_K \}$, and we denote $\z_0 = \x_{t+1}$ for convenience. This slight deviation from the formal definition of Monte Carlo integration is justified in this context for three reasons: (1) the resulting formulas have clear links to those presented in the literature, (2) it improves numerical stability and decreases bias for small $K$ (as illustrated with simulations in Appendix A), and (3) the effect of this change vanishes for large values of $K$. The approximate likelihood of a step under Monte Carlo integration is then
\begin{equation*}
    p(\x_{t+1} \mid \x_{1:t}) \approx \frac{w(\x_t, \x_{t+1}) \phi(\x_{t+1} \mid \x_{1:t})}{\frac{1}{K + 1} \sum_{k = 0}^K g(\z_k)},\quad \text{where } \z_k \sim f,
\end{equation*}
for some general $f$ and $g$.  We present several important special cases below, which encompass most existing methods for step selection analysis \citep[e.g.,][]{fortin2005}, including extensions proposed by \cite{forester2009}, \cite{duchesne2015}, and \cite{avgar2016}. A key characteristic of the first approach, described in Section \ref{sec:mc-fortin}, is that it assumes that the movement kernel $\phi$ is known prior to the step selection analysis. It has become increasingly common to jointly estimate the movement kernel and habitat selection, and the approaches in Sections \ref{sec:mc-unif}--\ref{sec:imp} focus on that situation.

\subsubsection{Assuming that the movement kernel is known}
\label{sec:mc-fortin}

A popular approach to step selection analysis is to define the movement kernel prior to fitting the SSF, typically from empirical or parametric distributions of step lengths and turning angles \citep{fortin2005}. Then, since $\phi$ is assumed to be known, we can apply Monte Carlo integration to the integral of Equation \ref{eqn:integral} by choosing $f = \phi$ and $g = w$. Equation \ref{eqn:integral} becomes
\begin{equation*}
  I \approx \frac{1}{K+1} \sum_{k = 0}^K w(\x_t, \z_k),\quad \text{where } \z_k \sim \phi(\cdot \mid \x_{1:t}).
\end{equation*}

That is, we generate the random Monte Carlo sample from the movement kernel $\phi$, and we take the mean of the habitat selection function at those random points to evaluate the integral. This characteristic of the typical step selection analysis workflow has led practitioners to view the $\z_k$ as a sample from the ``available'' landscape. The key limitation of this approach is that the movement kernel cannot be estimated jointly with habitat selection parameters, because the points $\z_k$ are sampled from it prior to model fitting. In addition, this approach to estimating $\phi$ without consideration of $w$ has been shown to result in biased parameter estimators since the observed movements are a function of both processes \citep{forester2009}.

Using this method, the likelihood of a step (Equation \ref{eqn:ssf}) is approximated by
\begin{equation}
  \label{eqn:ssf-mc}
  p(\x_{t+1} \mid \x_{1:t})
  \approx \frac{w(\x_t, \x_{t+1}) \phi(\x_{t+1} \mid \x_{1:t})}{\frac{1}{K+1} \sum_{k = 0}^K w(\x_t, \z_k)}
  \propto \frac{w(\x_t, \x_{t+1})}{\sum_{k = 0}^K w(\x_t, \z_k)},
\end{equation}
where $\z_k \sim \phi(\cdot \mid \x_{1:t})$. Note that multiplicative constants, i.e., terms that do not depend on the parameters $\bbeta$, can be omitted from the likelihood function with no effect on inference. Here, $\phi$ can be omitted from the numerator because it is assumed known, and therefore does not depend on any estimated parameter, and $1/(K+1)$ is omitted from the denominator. When $w$ is written in the usual exponential form, this is the likelihood of a conditional logistic regression (CLR) model, and parameters can be estimated using standard software such as the \texttt{clogit} function in the survival R package \citep{therneau2023}.  However, the method of Equation \ref{eqn:ssf-mc} has also been used for non-exponential functional forms for $w$, by using likelihood maximisation procedures other than CLR \citep{potts2014}.

This approach was initially proposed by \cite{fortin2005}, and has been widely used due to the convenience of implementation using CLR, and the intuitive appeal of interpreting the random locations as a sample of availability. Estimating the movement kernel separately has drawbacks, however: as previously mentioned, this approach leads to biased parameter estimators and  does not propagate statistical uncertainty about the movement model to the habitat selection parameters. In addition, it does not allow for interactions between the movement parameters and local environmental features. As a consequence, recent research in step selection analysis has focused on formulations where $\phi$ and $w$ are estimated simultaneously \citep{rhodes2005, forester2009, avgar2016}, and we present those in the next subsections. The interpretation of the random points $\z_k$ is different when $\phi$ is estimated as part of the SSF, and we discuss this in Section \ref{sec:rethinking}.

\subsubsection{Uniform Monte Carlo sampling}
\label{sec:mc-unif}

\begin{figure}[htbp]
  \centering
  \includegraphics[width=0.7\textwidth]{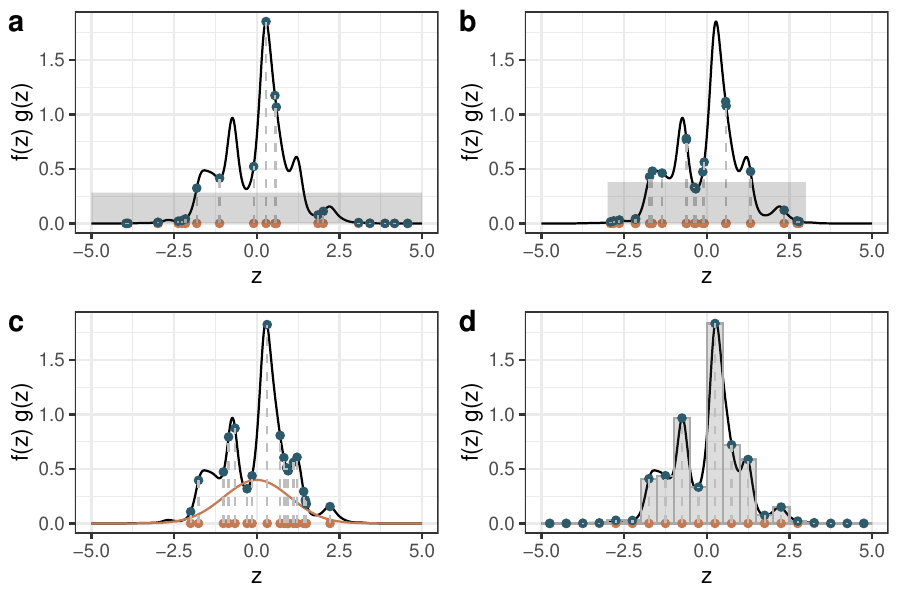}
  \caption{Illustration of numerical integration in one dimension, for the function $f \times g$ shown as a black line, over the interval $\Omega = [-5, 5]$. The orange dots are the integration points, and the blue dots are the corresponding function evaluations. (a) Monte Carlo integration with a uniform sample over $\Omega$; the height of the grey rectangle is the mean of function evaluations. (b) Monte Carlo integration with a uniform sample over $[-3, 3]$. (c) Importance sampling with random points generated from a normal distribution that roughly approximates the integrand $f \times g$. (d) Quadrature over a regular grid using a Riemann sum. In (a), (b) and (d), the shaded area approximates the area under the curve; there is no such simple visualisation method for importance sampling. In this small simulated example, the true integral is 2.53, and the approximations are (a) 2.83, (b) 2.28, (c) 2.41, and (d) 2.80.}
  \label{fig:integrals}
\end{figure}

If the movement kernel $\phi$ is not known a priori, we must choose a different probability density function $f$ from which to generate random points. One natural choice is to use a uniform distribution over the domain $\Omega$, because it is easy to sample from. In this case, the two functions in Equation \ref{eqn:mc1} are defined as $f(\z) = 1/A(\Omega)$ (where $A(\Omega)$ is the area of $\Omega$), and $g(\z) = A(\Omega) w(\x_t, \z) \phi(\z \mid \x_{1:t})$. Then, Equation \ref{eqn:integral} can be approximated as
\begin{equation*}
  I \approx \frac{A(\Omega)}{K+1} \sum_{k = 0}^K w(\x_t, \z_k) \phi(\z_k \mid \x_{1:t}),\quad \text{where } \z_k \sim \text{Unif}(\Omega).
\end{equation*}
Intuitively, the random points are used to estimate the mean value of the SSF over $\Omega$, and the integral is approximated by the product of that mean value and $A(\Omega)$ (see Figure \ref{fig:integrals}a for a one-dimensional example). 

This method, called ``uniform sampling" by \cite{forester2009}, allows for joint estimation of habitat selection and movement parameters \citep[e.g.,][]{schlagel2014detecting}. However, the uniform sampling approach can be  computationally demanding, because good performance requires that the integration points provide adequate coverage of the study region $\Omega$, and this can often only be achieved for large values of $K$. In most step selection analyses, the number of points can be greatly reduced based on the observation that, at each step, the SSF decreases sharply with distance from the start point $\x_t$ (due to movement constraints of the animal). Points far from $\x_t$ therefore contribute a negligible amount to the integral, and the approximation is virtually unchanged if the domain of integration is truncated to a disc around $\x_t$, with radius large enough to encompass any possible step \citep{boyce2003, craiu2008}. Fewer points are needed to ensure good coverage of this disc, which reduces the computational cost of evaluating the integral \citep{klappstein2022esf}. Uniform sampling on a truncated interval is illustrated in Figure \ref{fig:integrals}b.

Using this approximation, Equation \ref{eqn:ssf} becomes
\begin{equation}
\label{eqn:ssf-unif}
  p(\x_{t+1} \mid \x_{1:t})
  \approx \frac{w(\x_t, \x_{t+1}) \phi(\x_{t+1} \mid \x_{1:t})}{\frac{A(\Omega)}{K+1} \sum_{k = 0}^K w(\x_t, \z_k) \phi(\z_k \mid \x_{1:t})}
  \propto \frac{w(\x_t, \x_{t+1}) \phi(\x_{t+1} \mid \x_{1:t})}{\sum_{k = 0}^K w(\x_t, \z_k) \phi(\z_k \mid \x_{1:t})},
\end{equation}
where $z_k \sim \text{Unif}(\Omega)$. Like before, we remove $A(\Omega)/(K+1)$ from the denominator because it is a constant and thus will not affect maximum likelihood estimation. Here, we cannot omit $\phi$ from the numerator because it is not assumed to be known, and therefore is a function of the parameters $\bbeta$ of interest. If both $\phi$ and $w$ have an exponential form, then Equation \ref{eqn:ssf-unif} is equal to the CLR likelihood, as long as we include the observed location as an integration point. 

Note that uniform Monte Carlo sampling refers to \textit{spatially uniform points}, and this should not be confused with sampling points with uniform distances from $\x_t$ \citep{avgar2016}. Generating uniform distances will not result in a spatially uniform distribution of end points $\z_k$, and so the above formulas do not hold in that case. This is due to the fact that the set of possible long steps is spread over a larger area than the set of possible short steps; therefore if distances are uniform, points will be relatively more concentrated around the origin than far from it \citep{rhodes2005}. The case where distances are sampled from uniform distributions can in fact be viewed as a special case of importance sampling (Section \ref{sec:imp}).

\subsubsection{Importance sampling}
\label{sec:imp}

The precision of numerical integration depends on the choice of integration points; generally, the variability in the approximation is lower if points are concentrated in areas where the function takes large values. Importance sampling is a method to increase the precision of Monte Carlo integration by generating random points from a user-defined distribution $h$, called the importance function \citep[Section 3.3 of][and see an illustration in Figure \ref{fig:integrals}c]{robert2010}. To apply importance sampling to an SSF, we choose $f = h$ and $g = (w \times \phi) / h$ (so that $f \times g = w \times \phi$ as required), and Equation \ref{eqn:mc1} gives us the following approximation for the SSF integral,
\begin{equation}
\label{eqn:imp}
  I \approx \frac{1}{K+1} \sum_{k = 0}^K \frac{ w(\x_t, \z_k) \phi(\z_k \mid \x_{1:t})}{h(\z_k \mid \x_{1:t})},\quad \text{where } \z_k \sim h(\cdot \mid \x_{1:t}).
\end{equation}

The only constraint on $h$ is that it should be strictly positive over $\Omega$; when this is not the case, the approximation of the integral is truncated to the support of $h$ (i.e., the geographical area over which $h > 0$).  Note that, in Equation \ref{eqn:imp}, the importance function $h$ is used to weigh the contribution of each sampled point to the approximation; this is required to correct for the preferential sampling of some points over others when generating $\z_k$.

Importance sampling is useful because the function $h$ can be chosen in such a way that the variance of the integral estimator decreases (i.e., its precision increases). The aim is to choose a function $h$ with a shape that is as similar as possible to $g$ \citep[Section 6.6 of][]{rizzo2019}. This is a convenient framework in the context of SSFs, because it is often possible to determine where the SSF will take large values, based on the movement constraints of the animal. Animals are likely to avoid long steps, and so the SSF often decays rapidly as distance from the start point increases (Figure \ref{fig:ssf}). The speed of this decay can be determined approximately from the data, for example by fitting a distribution to the observed step lengths, and this information can be used to define the importance function $h$. For example, $h$ could be chosen as a bivariate normal distribution centred on the last location $\x_t$, or the two-dimensional spatial distribution implied by step length and turn angle distributions estimated from the data.

Using importance sampling with function $h$, the approximate likelihood of a step under the SSF model is
\begin{align}
    \label{eqn:ssf-imp}
    p(\x_{t+1} \mid \x_{1:t})
    & \approx \frac{ w(\x_t, \x_{t+1}) \phi(\x_{t+1} \mid \x_{1:t}) }{ \frac{1}{K+1} \sum_{k = 0}^K w(\x_t, \z_k) \phi(\z_k \mid \x_{1:t})/h(\z_k \mid \x_{1:t})} \nonumber \\
    & \propto \frac{ w(\x_t, \x_{t+1}) \phi(\x_{t+1} \mid \x_{1:t}) }{ \sum_{k = 0}^K  w(\x_t, \z_k) \phi(\z_k \mid \x_{1:t}) / h(\z_k \mid \x_{1:t})},
\end{align}
where $z_k \sim h$. The two other Monte Carlo approaches (Sections \ref{sec:mc-fortin}--\ref{sec:mc-unif}) can be viewed as special cases of importance sampling. If $h$ is the probability density function of a uniform distribution over geographical space, it is constant and can be omitted in Equation \ref{eqn:ssf-imp}, and we obtain Equation \ref{eqn:ssf-unif}. Alternatively, if $\phi$ is assumed to be known, and we choose $h = \phi$, then $h$ and $\phi$ cancel out in the denominator of Equation \ref{eqn:ssf-imp}; omitting $\phi$ from the numerator because is is known, this simplifies to Equation \ref{eqn:ssf-mc}.

Although the term ``importance sampling'' has rarely been used in the SSF literature, this and equivalent methods have been widely advocated, starting with the recommendation of \cite{forester2009} to distinguish between the sampling function $h$ and the movement model $\phi$. \cite{forester2009} derived a formula very similar to Equation \ref{eqn:ssf-imp}, but with one small difference: their numerator is divided by $h(\x_{t+1} \mid \x_{1:t})$. This difference is inconsequential because $h(\x_{t+1} \mid \x_{1:t})$ does not depend on the estimated parameters, so excluding it does not affect inference. The widely-used methods of \cite{avgar2016}, often called integrated step selection analysis, are based on \cite{forester2009} and can also be viewed as importance sampling. They focus on cases where $h$ and $\phi$ are chosen to be from the same exponential family of distributions, so that the calculations simplify and CLR can be used, but there is no such restriction when the approach is implemented with maximum likelihood estimation. In the approach of \cite{avgar2016}, the parameter estimates are adjusted after model fitting to correct for the sampling design. Here, the bias is corrected directly by including $h$ in Equation \ref{eqn:ssf-imp} when implementing the likelihood function. Another notable example is \cite{johnson2008}, who explicitly suggested importance sampling for a weighted distribution model analogous to the SSF in Equation \ref{eqn:ssf}, and used maximum likelihood estimation to fit the model without the need for CLR. More recently, \cite{klappstein2023} and \cite{pohle2023} also recognised that the approach proposed by \cite{forester2009} was a form of importance sampling.

\begin{figure}[htbp]
    \centering
    \includegraphics[width=0.9\textwidth]{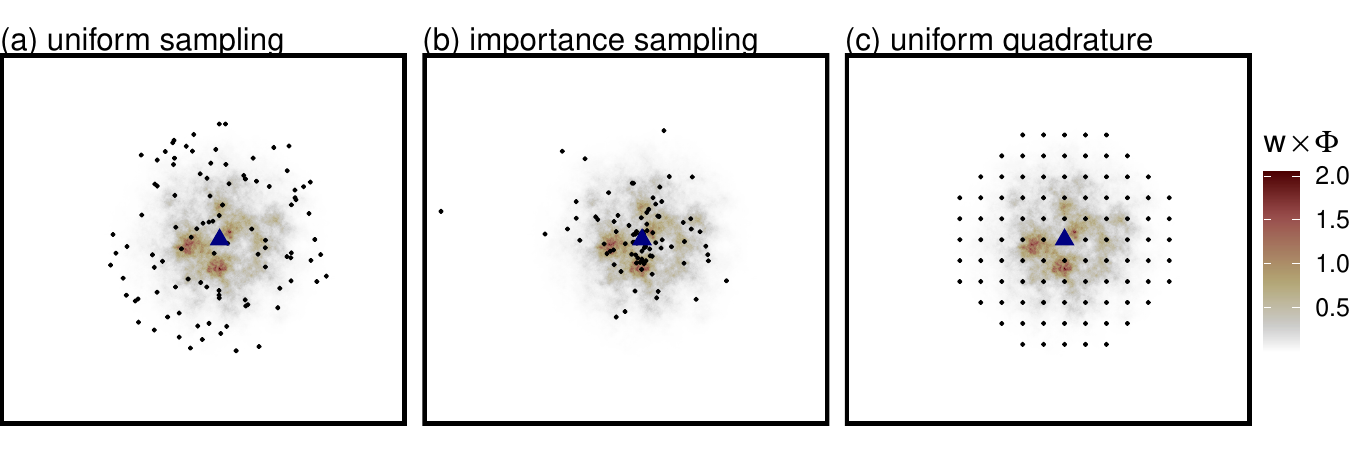}
    \caption{Illustration of three integration designs for an example SSF. The triangle in the centre shows the last location $\x_t$, and the heatmap shows the function that needs to be integrated, i.e., $w(\x_t, \z) \phi(\z \mid \x_{1:t})$. This function decreases with distance to $\x_t$ due to the animal's movement constraints. The black dots in each panel represent  100 points used for numerical integration: (a) uniform points over a disc, (b) points generated from an importance function based on distance to $\x_t$, and (c) regular quadrature grid over a disc. Importance sampling generates more points in areas where the function is high, and will typically lead to a better approximation of the integral.}
    \label{fig:ssf-samp}
\end{figure}

Figure \ref{fig:ssf-samp} contrasts importance sampling and uniform Monte Carlo sampling (Section \ref{sec:mc-unif}) for an example SSF. Similar to the one-dimensional example shown in Figure \ref{fig:integrals}, the intuition is that the precision of the approximation of the integral depends on the coverage of areas where the SSF takes high values. More specifically, the variance of the approximation is minimised when the distribution of integration points (i.e., the importance function $h$) is proportional to the SSF $w \times \phi$. In practice, $w$ and $\phi$ are not known, but a good heuristic is to approximate the movement kernel with parametric distributions (which can then be sampled from), and use these to define $h$.

\subsection{Quadrature}
\label{sec:quad}

Quadrature is a deterministic (non-random) alternative to Monte Carlo sampling, where the function is evaluated on a user-defined grid of points over the domain of integration. The simplest example of quadrature is the Riemann sum  (see Figure \ref{fig:integrals}d); in one dimension, the integral is approximated by the sum of the areas of rectangles with heights determined by function evaluations along a regular grid of points. The two-dimension extension of this approach consists of evaluating the function at regularly-spaced points along a two-dimensional grid, and approximating the integral by the sum of the volumes of cuboids.

In the context of an SSF, quadrature can be applied by evaluating $w(\x_t, \z_k) \phi(\z_k \mid \x_{1:t})$ at the centres of $K+1$ grid cells, $\{ \z_0, \z_1, \z_2, \dots, \z_K \}$, and calculating
\begin{equation*}
  I \approx A \sum_{k=0}^K w(\x_t, \z_k) \phi(\z_k \mid \x_{1:t}),
\end{equation*}
where $A$ is the area of each grid cell. Unlike Monte Carlo approaches, there is no variance in the estimate of the integral obtained from quadrature, because the grid is fixed (for a given $K$) rather than random. However, there is some error in the estimate (akin to statistical bias), and this error decreases as the spatial resolution of the quadrature grid increases (i.e., as $K$ increases).

\cite{rhodes2005} proposed this approach, and pointed out that the grid cells need not cover the entire study area, and the approximation can be truncated to the region where the function is non-negligible (within some distance of $\x_t$). This can speed up computations, similar to the observation in Section \ref{sec:mc-unif} that Monte Carlo integration with uniform sampling can be performed over a disc centred on the last location, as long as the radius of the disc is big enough. We call the approach where integration points are on a regular grid ``uniform quadrature'', and it is illustrated in Figure \ref{fig:ssf-samp}c for an example SSF. The equation for approximating the integral using uniform quadrature is identical to that obtained for uniform Monte Carlo sampling, and so the expression for the SSF likelihood is given by Equation \ref{eqn:ssf-unif}. 

In step selection analysis, environmental covariates are often available only over the discrete cells of a raster, making the centroids of the raster cells a natural choice for the grid of quadrature \citep{rhodes2005}, but this is not the only possible choice. \cite{arce2023} described a new method of inference for SSFs, where the function to integrate is evaluated at the nodes of a (deterministic) triangular mesh. This is another form of quadrature, which works well for the general class of spatial point processes implemented in the \texttt{inlabru} spatial modelling software that they use \citep{simpson2016, bachl2019}. 

Some have argued that deterministic numerical integration is preferable in habitat selection modelling because it has better properties than Monte Carlo integration for low-dimensional integrals like the one in Equation \ref{eqn:integral} \citep{warton2010, arce2023}. However, this comparison assumed uniform Monte Carlo sampling, for which the performance can be improved substantially using importance sampling with a well-chosen function $h$. An interesting alternative might therefore be adaptive quadrature, where the spatial arrangement of points in a (deterministic) spatial grid is chosen based on the shape of the function. The general idea is to iteratively subdivide the domain of integration, in such a way that regions where the function is more irregular are subdivided further. This is analogous to the idea in importance sampling of generating points in regions where the function is more complex or takes higher values \citep{pinheiro2006}.

\section{Illustration}

We illustrate some of the key concepts and methods using simulations and a real data analysis. The general approach to fitting SSFs presented in Section \ref{sec:mle} requires implementing the (approximate) likelihood function, rather than relying on existing CLR software. Writing custom code greatly increases the flexibility of the model formulation and inference methods. To help readers implement their own step selection analyses, we provide R code that can be used as a starting point. We aimed for a trade-off between simplicity and flexibility, and we provide basic functions that can be tailored to fit a wide range of model formulations. The documented code and detailed examples are provided in Appendix C.

\subsection{Comparing sampling designs in simulations}
\label{sec:design}

Different methods of fitting SSFs can be viewed as different numerical integration approaches for the same underlying model. When using Monte Carlo and quadrature methods, the placement of integration points is known to affect the precision of the results \citep{rizzo2019}, and this provides a rigorous grounding for the intuition that some methods of generating locations perform better than others. The closer the distribution of integration points is to the true SSF, the more precise the approximation will be. In this section, we use simulations to assess the effect of several design choices on our ability to recover the parameters of an SSF. We do not intend these simulations to be exhaustive, or to provide general guidelines to select the best sampling design in every application, as the choice of method and model is study-specific. Rather, our aim is to showcase the qualitative effect of different design choices, encouraging researchers to explore the various options for themselves when they perform step selection analysis .

We considered three designs for the integration points: (1) points sampled uniformly at random over a disc of radius equal to the maximum observed step length, (2) importance sampling where $h$ was based on a gamma distribution of distances and a von Mises distribution of turning angles, and (3) uniform quadrature where integration points were defined on a regular spatial grid over a disc of radius equal to the maximum observed step length. In all cases, the parameters of the importance function were chosen based on the empirical distributions of step lengths and/or turning angles.

We first simulated a movement track of length 1000 from an SSF with known parameters. For the habitat selection component, we used $w(\x_t, \x_{t+1}) = \exp(\beta \times c_1(\x_{t+1}))$, where $c_1$ was a simulated spatial covariate shown in Appendix B and $\beta = 5$ was the corresponding selection parameter. The movement kernel was chosen as $\phi(\x_{t+1} \mid \x_{1:t}) = \exp(-4 L_t + 3 \cos(\alpha_t))$, where $L_t$ and $\alpha_t$ are the step length and turning angle at time $t$, respectively. This formulation implies that step lengths followed a gamma distribution with mean 0.5 and standard deviation 0.35, and turning angles followed a von Mises distribution with mean zero and concentration 3. For each scenario, we ran the following steps for increasing numbers of random points, $K \in \{ 5, 10, \dots, 195, 200 \}$:
\begin{enumerate}
    \item Generate $K$ integration points for each observed step, based on a given numerical integration method;
    \item Estimate the selection parameter $\beta$ using maximum likelihood estimation, based on the integration points from step 1.
\end{enumerate}
For Monte Carlo sampling methods, we repeated these steps 50 times to capture estimator variability, yielding 50 estimated parameters $\widehat{\beta}$ for each approach and each number of random points. For deterministic quadrature, there is no sampling variability and so the procedure only needed to be run once for any given $K$. Note that, for quadrature, the number of integration points was constrained by the design of the grid, and so it does not always exactly match the values of $K$ listed above. For each simulation, we evaluated estimator bias due to the numerical integration, as the difference between the estimated parameter and the asymptotic estimate obtained when $K$ is very large (here, $K = 5000$). We did not compare the estimates to the true parameter values used in simulation, because this would not make it possible to separate bias due to integration error (which we are interested in) and finite-sample bias inherent to maximum likelihood estimation (which does not depend on integration method).

\begin{figure}[htbp]
    \centering
    \includegraphics[width=\textwidth]{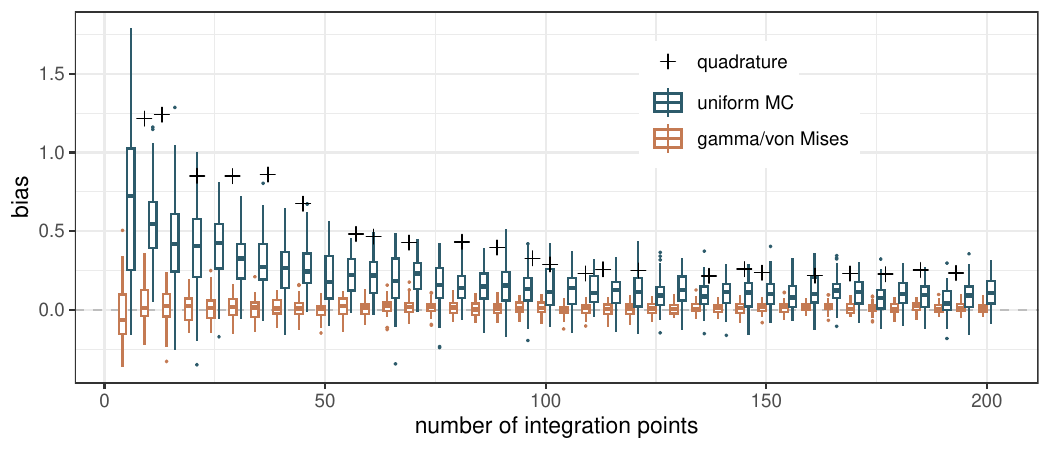}
    \caption{Results of simulation study. Bias (estimate $-$ truth) in estimated selection parameter for one spatial covariate, as function of the number of integration points. Results are compared for three methods for performing numerical integration: uniform Monte Carlo (blue boxes), importance sampling with gamma-distributed distances and von Mises-distributed angles (red boxes), and uniform quadrature (black crosses). Each box represents 50 estimated values. For quadrature, the number of integration points was constrained by the design of the grid, so values of $K$ do not exactly match those used for other methods.}
    \label{fig:sim}
\end{figure}

Figure \ref{fig:sim} compares the bias in the selection parameter $\beta$ for the three numerical integration approaches, over a range of numbers of random points. The bias and variance decreased as the number of random points increased for all tested methods. Overall, uniform Monte Carlo sampling had lower precision (i.e., higher variation) than importance sampling; this is to be expected, as importance sampling uses observed movement patterns to generate integration points efficiently. Uniform quadrature was generally positively biased (i.e., the selection parameter was overestimated) and, although the bias decreased as $K$ increased, both random sampling methods seemed to perform better in this analysis. As described in Section \ref{sec:mc}, the bias is due to the log-transformation of the likelihood for optimisation: the parameter value that maximises the approximate log-likelihood can be biased even if the estimator of the likelihood is unbiased \citep{gourieroux1996}.

These simulations show that the performance of any method of numerical integration is dependent on the number of integration points. In practice, there is no consensus about how high this number should be, and \cite{thurfjell2014} reported that different studies have used a wide range of numbers of random points, between $K = 2$ and $K = 200$. The minimum number needed for a given analysis depends on many factors, such as the length of the observed times series, the sampling scheme used, and the complexity of the SSF model formulation. For this reason, we recommend that practitioners try several numbers of random points, until the parameter estimates stabilise, to ensure that the approximation error in the results is small. This is consistent with the results presented in Figure 2 of \cite{fieberg2021}, and similar to the advice of \cite{warton2010} and \cite{northrup2013} in the context of resource selection functions. Small samples can also lead to numerical instability and failure to converge during model fitting.

\subsection{Comparing models using AIC and the same set of integration points}
\label{sec:applied}

To demonstrate how an understanding of numerical integration techniques and their use in step selection analyses can facilitate model comparisons, we considered location data from a red deer (\textit{Cervus elaphus}) in Northern Germany, automatically loaded with the amt R package as the data object \texttt{deer} \citep{signer2019}. The locations are on a regular 6-hour time grid, and the package also provides a binary raster layer for forest cover (through the \texttt{get\_sh\_forest()} function). We will use importance sampling with a single set of integration points to fit multiple models and compare them using AIC. 

We generated integration points using gamma-distributed step lengths and von Mises-distributed turning angles, with parameters estimated from their empirical distributions. We compared four SSF formulations, all with the same habitat selection component $w$ (with forest cover as covariate), but with different families of distributions of step lengths $L_t$ and turning angles $\alpha_t$ in the movement kernel: (i) $L_t \sim \text{Exp}(\theta_1)$ and $\alpha_t \sim \text{uniform}(-\pi, \pi)$, (ii) $L_t \sim \text{gamma}(\theta_1, \theta_2)$ and $\alpha_t \sim \text{uniform}(-\pi, \pi)$, (iii) $L_t \sim \text{gamma}(\theta_1, \theta_2)$ and $\alpha_t \sim \text{von Mises}(0, \kappa)$, and (iv) $L_t \sim \text{Weibull}(\theta_1, \theta_2)$ and $\alpha_t \sim \text{wrapped Cauchy}(0, \kappa)$. Note that we make a distinction between the distributions used to generate random locations (which determine the importance function $h$) and the distributions  used to specify the movement kernel $\phi$. This allows us to fit all four SSFs using the same integration points, i.e., the exact same data set, such that the models can be compared using AIC. 

The results are shown in Table \ref{tab:AIC}. For this data set, AIC favoured the SSF formulation (iv), where step lengths were modelled with a Weibull distribution and turning angles with a wrapped Cauchy distribution. This example illustrates two advantages of implementing maximum likelihood estimation using numerical integration: the flexibility of choosing non-exponential models for the movement kernel (the Weibull and wrapped Cauchy distributions are not in the exponential family), and the ability to compare models with different movement formulations. 

\begin{table}[htbp]
    \centering
    \begin{tabular}{lllcc}
        \toprule
         & Step length & Turning angle & AIC & $\Delta$AIC  \\
         \midrule
         (iv) & Weibull & wrapped Cauchy & 2500 & 0 \\
         (ii) & gamma & uniform & 2507 & 7 \\
         (iii) & gamma & von Mises & 2509 & 9 \\
         (i) & exponential & uniform & 2521 & 21 \\
         \bottomrule
    \end{tabular}
    \caption{Model comparison for deer analysis. The four SSFs are specified using different distributions of step lengths and turning angles, and they are shown in order of increasing AIC (i.e., the better model is at the top). $\Delta$AIC is the difference in AIC between each model and the better model.}
    \label{tab:AIC}
\end{table}

\section{Rethinking step selection analysis}
\label{sec:rethinking}

\subsection{Beyond conditional logistic regression}

It is common for step selection analysis to be presented as conditional logistic regression (CLR), where each observed step is a ``case'' associated with a set of ``control'' (random) steps. Indeed, in many important special cases, the SSF likelihood is approximately equivalent to that of a CLR model \citep{fortin2005, forester2009}, which allows parameters to be estimated  with CLR software \citep{signer2019, therneau2023}. To justify that approach, \cite{forester2009} described an SSF as a discrete-choice model, which is a popular way to present resource selection functions \citep{cooper1999}. In that framework, the animal is assumed to have access to a finite number of discrete and mutually exclusive resource units, and the model describes how it chooses one unit over the others. In a step selection model, however, the animal has infinitely many movement options (because it moves over a continuous space). To reduce the problem to a discrete choice, \cite{forester2009} conditioned each movement decision on a set of random (``control'') points, and assumed that those represented the animal's options for that time step. This can be viewed as an approximation of the target model, akin to the numerical integration approaches that we presented in Section \ref{sec:mle}, and the two approaches lead to equivalent formulas.

The equivalence between CLR and the SSF likelihood has allowed ecologists to leverage standard statistical software for quantifying drivers of movement and habitat selection for many years. Recently, \cite{muff2020} and \cite{chatterjee2023} demonstrated how a similar equivalence between CLR and Poisson regression with fixed stratum-specific intercepts could be exploited to model individual variability in habitat selection and movement parameters using random effects.  Despite these equivalencies, we emphasise that CLR is not the model of interest, but rather a tool to fit the target SSF model, shown in Equation \ref{eqn:ssf}. Notably, the equivalence between CLR and the SSF likelihood only holds when both the habitat selection function $w$ and the movement kernel $\phi$ have an exponential form. CLR is therefore limited to distributions from the exponential family when modelling $\phi$, and most step selection modelling research has therefore defaulted to using an exponential or gamma distribution for step lengths and the von Mises distribution for turn angles \citep{duchesne2015, avgar2016}. There is no such restriction in the general model, however, which allows for the use of a much wider range of distributions, such as the Weibull distribution for step length and the wrapped Cauchy distribution for turning angle. As our applied example  illustrates (Section \ref{sec:applied}), these alternatives can potentially lead to improved model fit. Likewise, it is possible to model habitat selection using functions other than the exponential \citep[e.g.,][]{potts2014, schlagel2017}, although it is a natural choice for continuous variables since selection functions should be positive and unbounded \citep{mcdonald2013}.

% Although it is common to split SSFs into two functions as shown in Equation \ref{eqn:ssf}, this separation is not a necessity. The two functions can be combined if $\phi$ is written in an exponential form, i.e., $\phi(\x_{t+1} \mid \x_{1:t}) = \exp \{ m(\x_{1:t}, \x_{t+1}) \bbeta_m^\intercal \}$, where $m(\x_{1:t}, \x_{t+1})$ is a vector of movement variables, and $\bbeta_m$ a vector of movement parameters \citep{avgar2016, klappstein2023}. This expression for $\phi$ is fairly flexible and, for the right choice of covariate, it can for example capture a gamma distribution for step length and a von Mises distribution for turning angle \citep{avgar2016, duchesne2015}. \cite{avgar2016} argued that it is convenient to combine $w$ and $\phi$ as it makes it possible to include interactions between movement and habitat components of the model, and it allows for joint estimation of movement and habitat selection parameters within a conditional logistic regression framework. 

\subsection{Separating model and inference}

As we have already mentioned, it is important to distinguish between the choice of model formulation and the method of parameter estimation. Here, the model is of the form shown in Equation \ref{eqn:ssf}, and the main modelling decision is defining the functional forms of $w$ and $\phi$. For a given model formulation, many possible numerical integration methods can be used to approximate the likelihood, and therefore to estimate the model parameters, as described in Section \ref{sec:mle}. These methods, together with any implementation scheme such as CLR or other likelihood maximisation, constitute the inference procedure, and are not the model itself. Choices of model and inference procedure are both important for the analysis: the model formulation should capture important features of the data-generating process (i.e., animal movement and habitat selection), whilst the inference procedure should be chosen carefully to reduce the approximation error. We posit that much confusion has arisen in the context of step selection analysis due to the conflation of model and inference, especially when comparing techniques.

One particular area of confusion has been the interpretation of the integration points $\{ \z_1, \dots, \z_K \}$ generated as part of model fitting. Due to the historical influence of \cite{fortin2005}, who assumed that the movement model $\phi$ was known and used it to generate random locations (Section \ref{sec:mc-fortin}), the $\z_k$ are commonly assumed to be a sample of ``available'' points (or, equivalently, steps connecting the previous location and $\z_k$ are seen as possible movements). However, this is crucially not the correct interpretation in other approaches, such as uniform sampling (Section \ref{sec:mc-unif}) and importance sampling (Section \ref{sec:imp}). Indeed, with those methods, the sample of random locations does not have any particular biological interpretation, and it merely constitutes a numerical tool to approximate an integral over space. This is the case in most modern step selection analyses, in which availability is estimated jointly with habitat selection through the parameters of the function $\phi$, rather than assumed known a priori \citep{rhodes2005, forester2009, avgar2016}.

Another important distinction is between the distributions used to model movement, and the distributions used to generate random locations. The modelled distributions are specified through the choice of $\phi$, independently of the choice of sampling function (e.g., $h$ in importance sampling). \cite{forester2009} and \cite{avgar2016} showed that it is mathematically convenient to generate random locations from the same family of distributions that is used in $\phi$, but this is not a necessity. As we demonstrate in Section \ref{sec:applied}, it is for example possible to use a gamma distribution of distances to generate random locations (i.e., in the importance function $h$), but specify $\phi$ so as to model step lengths with a Weibull distribution in the analysis. Furthermore, it is straightforward to fit SSFs with different movement kernel formulations on the exact same data set (including identical random locations). This makes it possible to select the formulation for the movement kernel based on standard model selection criteria such as AIC.  This is not possible within the workflow outlined by \cite{avgar2016} and implemented by \cite{signer2019}, which requires $h$ and $\phi$ to be from the same family. When the model and method of inference are separate, it is also easier to determine how different functional forms for the movement kernel $\phi$ can be implemented in practice, and we discuss this in Appendix D.

In fact, the distribution from which the random locations are generated only matters insofar as different distributions might require different numbers of integration points to achieve low error (Section \ref{sec:design}). For a large enough number of integration points, the choice of distribution is inconsequential. In practice, when the computational cost is high, it might not be an option to increase the size $K$ of the random sample arbitrarily; in this case, it is preferable to choose an importance function $h$ that reduces the estimation variance for a given $K$. Importantly, the choice of $h$ does not reflect a modelling assumption, and it is used merely as an inferential tool to reduce the approximation error.

\section{Conclusion}

The general problem of evaluating complex integrals has been studied extensively, and many different approaches could be adapted in the context of  step selection analysis. For example, the Laplace approximation is a versatile method where it is not necessary to evaluate the function at integration points. Instead, it is assumed that the integrand is well approximated by a normal distribution, for which the integral is known. In step selection analyses, the integrand typically combines the movement kernel and the habitat selection function, resulting in a complex (possibly multimodal) function, and it is not clear whether the Gaussian assumption would be reasonable. As suggested in Section \ref{sec:quad}, another promising direction is to combine ideas from quadrature and importance sampling, and choose the nodes of a deterministic grid to improve the approximation in areas where the SSF is irregular. This is similar to adaptive quadrature, which is already widely used in ecology for non-Gaussian generalised linear mixed models \citep{bolker2009}.

All model fitting approaches are approximately equivalent in the limit where the number of points in the Monte Carlo sample or the quadrature grid is large (i.e., as $K \to \infty$). It might therefore seem unnecessary to concern ourselves with the design choices described in this review, because we can always increase $K$ until the bias and/or variance in the estimation are negligible. Because SSFs are relatively simple models, computation time is often small for modest-sized data sets, and the additional cost of increasing $K$ might be moderate. However, we have shown that using simplistic numerical integration techniques can cause bias to persist even for quite large $K$ (e.g., uniform Monte Carlo and quadrature; Figure \ref{fig:sim}), so thinking carefully about the numerical integration technique employed may often be preferable to simply increasing $K$.  Furthermore, step selection models are becoming more sophisticated and complex, so computational effort might become the bottleneck of many analyses \citep[e.g., multistate models;][]{nicosia2017}. We anticipate that the sampling design will become increasingly critical in those cases, as well as in studies with large data sets.

We are certainly not recommending that all biologists stop using CLR software for step selection analyses, as it is a convenient, fast, and stable implementation for many purposes. In cases where the habitat selection and movement are modelled in exponential form, the CLR approach outlined by \cite{forester2009}, \cite{duchesne2015}, and \cite{avgar2016} can safely be applied, e.g., with the amt R package \citep{signer2019}. However, even in that context, understanding the role numerical integration plays in parameter estimation can shed light on several important technical details. In particular, this approach clarifies the role of ``control'' points in step selection analysis: although those locations are generally a numerical tool rather than a biologically-relevant set of spatial locations, placing them in areas where the animal is most likely to move decreases the required number of points. In addition, importance sampling explains the post-hoc parameter adjustment proposed by \cite{avgar2016}, which is needed to correct for the choice of importance function (see also Appendix S3 of that paper). Overall, we have shown that the interpretation of step selection models ought to be separated from any specific model fitting approach, and that understanding this dichotomy between model construction and model fitting leads to much broader application of step selection functions than those confined to a CLR approach.

\subsection*{Acknowledgements}

JF was supported by National Aeronautics and Space Administration award 80NSSC21K1182 and received partial salary support from the Minnesota Agricultural Experimental Station.

\bibliographystyle{apalike}
\bibliography{refs.bib}

\end{document}